# Phenomenology of Large Extra Dimensions Models at Hadrons Colliders using Monte Carlo Techniques
## (Spin-2 Graviton)

Nady Bakhet[1, 2], Maxim Yu Khlopov[3, 4], Tarek Hussein[1, 2]

[1]Department of Physics, Cairo University, Giza, Egypt

[2]Egyptian Network of High Energy Physics - ASRT, Cairo, Egypt

[3]APC Laboratory, IN2P3/CNRS, Paris, France

[4]National Research Nuclear University "MEPHI" (Moscow Engineering Physics Institute) Russia

## ABSTRACT

Large Extra Dimensions Models have been proposed to remove the hierarchy problem and give an explanation why the gravity is so much weaker than the other three forces. In this work, we present an analysis of Monte Carlo data events for new physics signatures of spin-2 Graviton in context of ADD model with total dimensions $D = 4 + \delta$, ($\delta = 1,2,3,4,5$ and $6$) where $\delta$ is the extra Spatial dimension, this model involves missing momentum $P_T^{miss}$ in association with jet in the final state via the process $pp(\bar{p}) \rightarrow G + jet$, Also, we present an analysis in context of the RS model with 5-dimensions via the process $pp(\bar{p}) \rightarrow G + jet$, $G \rightarrow e^+e^-$ with final state $e^+e^- + jet$. We used Monte Carlo event generator Pythia8 to produce efficient signal selection rules at the Large Hadron Collider with $\sqrt{s} = 14\ TeV$ and at the Tevatron $\sqrt{s} = 1.96\ TeV$.

# 1 Large Extra Dimensions Models

In different new physics Models at the TeV scale, new spin-2 particles and in particular Gravitons are predicted. Today the search for new physics such as Graviton at the TeV scale is among the main tasks of the Large Hadron Colliders (LHC). The Large Extra Dimensions have been proposed to remove the hierarchy problem [1] and to give reasons why the gravity is weaker than the other three forces. There are many reasons to expect new physics beyond the SM appearing at the LHC TeV energy scale, such as the dark matter candidate and the problem on the large hierarchy between the electroweak scale and Planck scale. Different Extra Dimension models which have been proposed can be divided into two main classes according to the geometry of the background space-time manifold. The first one includes the ADD [2], which extend the dimension of the total space-time to D = 4+δ. The second one includes the 5-dimensional RS model [3, 4], in which a warped metric is introduced along the 5-th dimension the size of the extra dimension can be at the order of the Planck length.

In both scenarios of Extra Dimensions Models, Kaluza-Klein (KK) towers of massive spin-2 gravitons appear, that can interact with the Standard Model fields. The effective interaction Lagrangian is given by [5, 6]:

$$\mathcal{L}_{int} = \frac{1}{\Lambda} \sum_{\vec{n}} T^{(\vec{n})\mu\nu} \tau_{\mu\nu}$$

Where $T^{(\vec{n})\mu\nu}$ is the $\vec{n}th$ graviton KK modes, and $\Lambda$ is the relevant coupling scale.

## 2 ADD model at Hadron Colliders

The most signatures of ADD model at the hadron colliders involve missing momentum $P_T^{miss}$ in association with jet via the process $pp(\bar{p}) \rightarrow G + jet$. We use Monte Carlo event generator (MCEG) Pythia8 to produce efficient signal selection rules at the Large Hadron Collider (proton-proton collisions) with $\sqrt{s} = 14\,TeV$ and at the Tevatron (proton-antiproton collisions) with $\sqrt{s} = 1.96\,TeV$. We will use the G + jet channel to compare the ADD model results with the RS model results. We selected the events based on the presence of a jet with transverse energy $E_T > 100\,GeV$. The resulting candidate events contains a significant number of events originating from the Standard Model processes which can produce large missing transverse energy $\not{E}_T$. The $\not{E}_T$ associated with these processes can originate from neutrinos in the final state. The largest the Standard Model background is Z+ jets where the Z boson subsequently decays into neutrinos $Z \rightarrow \nu\bar{\nu}$ as shown in Figure 3, this is the main dominant Standard Model background in the ADD model for the process $pp(\bar{p}) \rightarrow G + jet$. This background has the same event topology as our signal and is thus irreducible. The next most significant Standard Model background comes from $w(\rightarrow \ell\nu) + jets$ production where $\ell = e$, $\mu$, or $\tau$ and the lepton is unidentified. The number of electroweak background events in the sample is estimated by the cross sections for $Z(\rightarrow \ell\ell) + jets$ and $W(\rightarrow \ell\nu) + jets$ ($\ell = e$ or $\mu$). We select events that contain $\mu$ with $p_T > 20\,GeV$ or an electron with $E_T > 25\,GeV$ using standard lepton selection rules [20] to establish a background sample of lepton candidates. Also there are additional selections to reject events (the noise and non-collision backgrounds), for the jet with $p_T > 20$ GeV and $|\eta| < 4.5$ does not pass all of the additional selection rules in [7]. Here the results are interpreted in the contexts ADD LED model and ADD model parameters, $M_D$ values are 5 TeV and 1 TeV at the LHC and the Tevatron respectively and a number of extra dimensions changes from 2 to 4. On the other hand, the Graviton emission is within reach of the Tevatron provided that the 4 + n dimensional Planck scale ($M_D$) is around 1 TeV [8, 9]. But for graviton production in the ADD scenario, a low-energy effective field theory [10] is used to calculate the signal cross section considering the contribution of different graviton mass modes. In this work, the Signal samples corresponding to a number of extra dimensions varying between 2 and 6 are considered. Also we use MadGraph5/Madevent [11–13], where a spin-2 particle implementation at the matrix element level [14]. Current terrestrial test of gravity set a limit on $M_D \geq 3.6\ TeV\ for\ \delta = 2$ [15]. Searches for the ADD graviton production have been performed in the processes $e^+e^- \rightarrow \gamma(z) + E^{miss}$ at LEP and $p\bar{p} \rightarrow \gamma(jet) + p_T^{miss}$ at the Tevatron. The combined LEP limits [16] read $M_D > 1.60, 1.20, 0.94, 0.77\ and\ 0.66\ TeV$ for $\delta = 2, ...,6$ respectively while Tevatron searches exclude $M_D > 1.40, 1.15, 1.04, 0.98, and\ 0.94\ TeV$ for $\delta =\ 2, ...,6$ respectively [17–19]. The SM background to the monojet signature is dominated by $Z \rightarrow \nu\bar{\nu}$ in RS model and $Z \rightarrow \ell^{\pm}\ell^{\mp}$ $(\ell = e, \mu)$ Drell-Yan process in ADD model.

## 2.1 Process $pp(\bar{p}) \rightarrow G + jet$

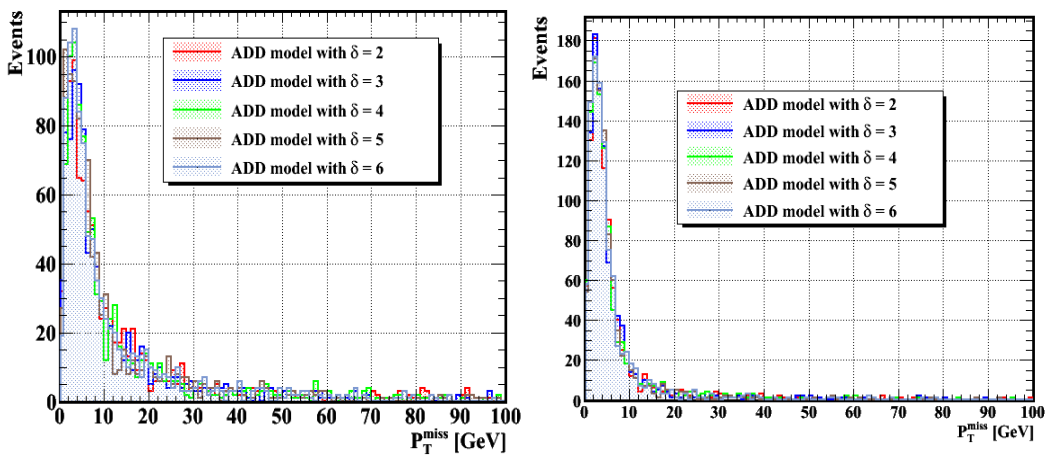

FIG. 1: Comparison for Graviton (missing transverse momentum) in the context of ADD model at the LHC $\sqrt{s} = 14\ TeV\ with\ \Lambda = 5\ TeV$ (LEFT) and at the Tevatron $\sqrt{s} = 1.96\ TeV$ with $\Lambda = 1\ TeV$ (RIGHT) $for\ \delta = 2,3,4,5\ and\ 6$ extra spatial dimensions.

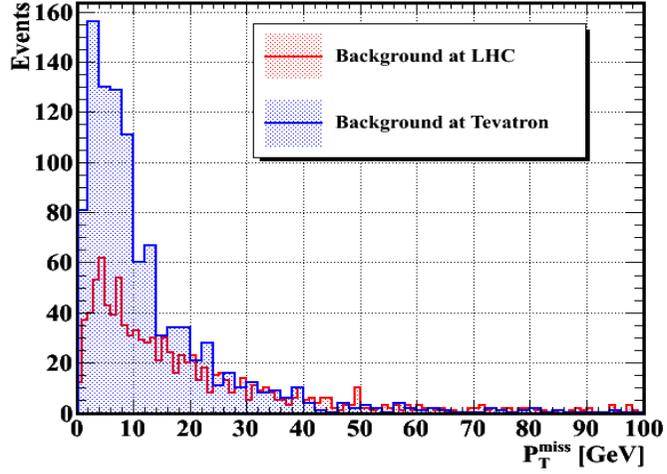

FIG. 2: The SM background missing transverse momentum $Z \to \nu\bar{\nu}$ of Graviton production in the context of ADD model at the LHC $\sqrt{s} = 14\,TeV$ and at the Tevatron $\sqrt{s} = 1.96\,TeV$.

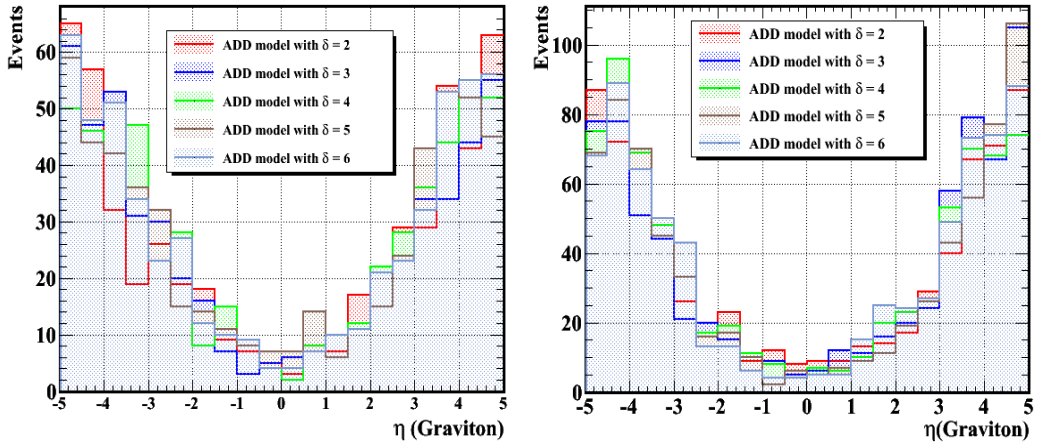

FIG. 3: Comparison for Graviton Pseudorapidity in the context of ADD model at the LHC $\sqrt{s} = 14\,TeV\ with\ \Lambda = 5\,TeV$ (LEFT) and at the Tevatron $\sqrt{s} = 1.96\,TeV$ with $\Lambda = 1\,TeV$ (RIGHT) $for\ \delta = 2,3,4,5\ and\ 6$ extra spatial dimensions.

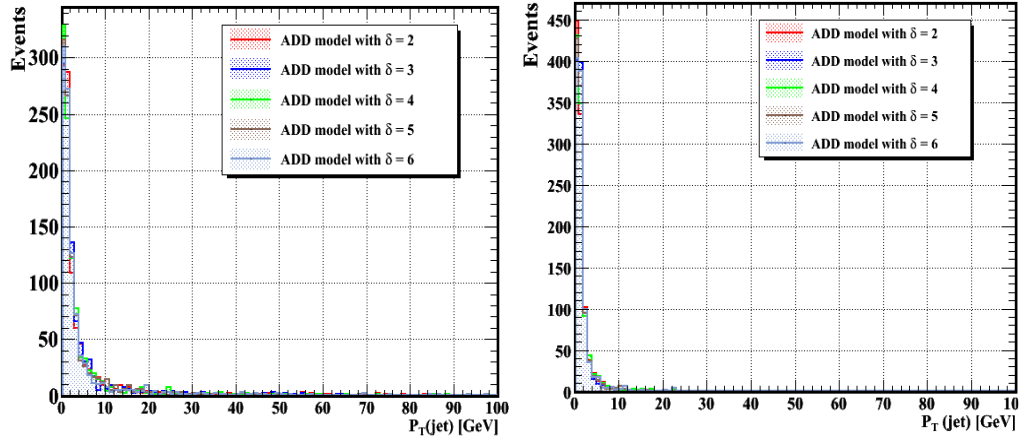

FIG. 4: Comparison for jet transverse momentum in the context of ADD model at the LHC $\sqrt{s} = 14\,TeV$ $with\,\Lambda = 5\,TeV$ (LEFT) and at the Tevatron $\sqrt{s} = 1.96\,TeV$ with $\Lambda = 1\,TeV$ (RIGHT) $for\,\delta = 2,3,4,5\,and\,6$ extra spatial dimensions.

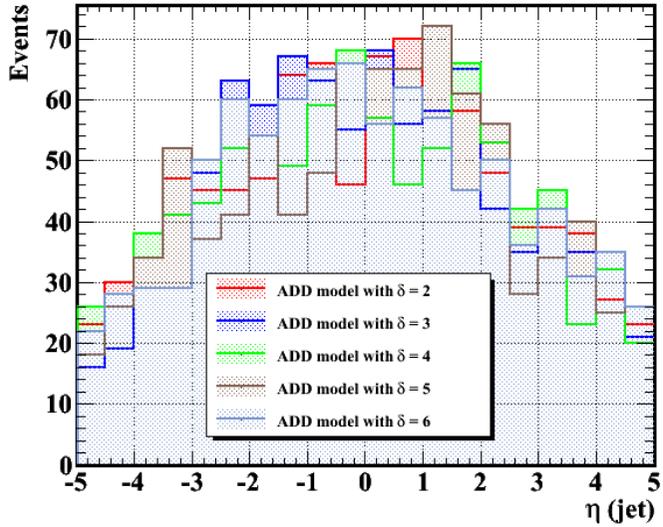

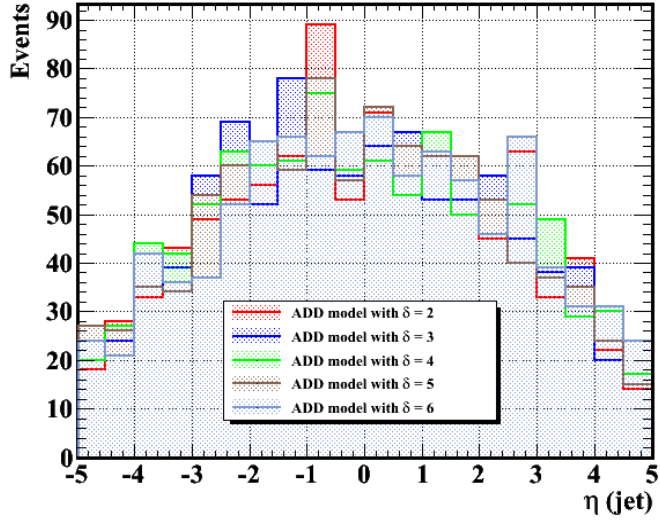

FIG. 5: Comparison for jet Pseudorapidity in the context of ADD model at the LHC $\sqrt{s} = 14\,TeV$ $with$ $\Lambda = 5\,TeV$ (TOP) and at the Tevatron $\sqrt{s} = 1.96$ TeV with $\Lambda = 1\,TeV$ (BOTTOM) for $\delta = 2,3,4,5\ and\ 6$ extra spatial dimensions.

## 3 RS model at Hadron Colliders

The phenomenology of (Randall Sundrum) RS model at the hadron colliders is Graviton spin-2 resonance in lepton and mono-jet production. The Graviton G decays to dielectron $e^+e^-$ via the process $pp(\bar{p}) \rightarrow G + jet$, $G \rightarrow e^+e^-$ and the final state will be $e^+e^-$ + jet. We will simulate this process using Monte Carlo programs at the Large Hadron Collider with $\sqrt{s} = 14$ TeV and at the Tevatron with $\sqrt{s} = 1.96$ TeV, so we will use the G + jet channel to compare the RS model results with the ADD results that we produced before. In the Standard Model, lepton pairs at hadron colliders can be produced at tree level via the parton-level processes: $q\bar{q} \rightarrow Z, \gamma \rightarrow \ell^+\ell^-$. The first massive graviton mode of the RS model, in the sequel denoted simply as G (and the mass $m_1 \equiv M_G$), can be produced via quark–antiquark annihilation as well as gluon–gluon fusion and can be observed as a peak in the dilepton invariant mass distribution. $q\bar{q} \rightarrow G \rightarrow \ell^+\ell^-$ and $gg \rightarrow G \rightarrow \ell^+\ell^-$. The selection rules for final state of leptons, for electrons candidate are required to have $p_T > 20\ GeV$ and $|\eta| < 2.47$, Muons candidates are required to have $p_T > 20\ GeV$ and $|\eta| < 2.4$. Muons are required to be isolated to reduce the background contribution from jet which consists of particles originating from a high $p_T$ jet [20]. Figure 6 shows the distribution of the invariant mass of the two electrons produce from Graviton decay process in the context of RS model at the LHC $\sqrt{s} = 14$ and 8 TeV with $\Lambda$ = 5 TeV and Graviton mass = 1 TeV and at the Tevatron $\sqrt{s} = 1.96$ TeV with $\Lambda = 1$ TeV and Graviton mass = 0.5 TeV. We reconstructed the Graviton mass at the LHC and the Tevatron for 1 TeV and 0.5 TeV respectively. There are two peaks at these two values. Also Figure 7 shows the distributions of the transverse momentum and pesudorapidity for the final state $e^+e^- + jet$ and Graviton in the context of RS model at the Large Hadron Colliders $\sqrt{s} = 14\ and\ 8\ TeV\ with\ \Lambda = 5\ TeV$ for Graviton mass = 1 TeV and at the Tevatron $\sqrt{s} = 1.96\ TeV$ with $\Lambda = 1\ TeV\ and\ m = 0.5\ TeV\ using\ Pythia8$.

## 3.1 Process $pp(\overline{p}) \rightarrow G(\rightarrow e^+e^-) + jet$

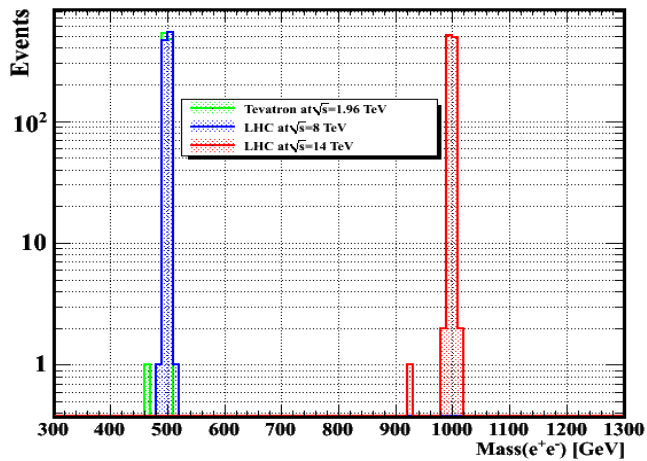

FIG. 6: Invariant mass of the two electrons produce from Graviton decay process in the context of RS model at the LHC $\sqrt{s} = 14\ and\ 8\ TeV\ with\ \Lambda = 5\ TeV\ and\ Graviton\ mass = 1\ TeV$ and at the Tevatron $\sqrt{s} = 1.96\ TeV$ with $\Lambda = 1\ TeV$ and Graviton mass equal 0.5 TeV using Pythia8

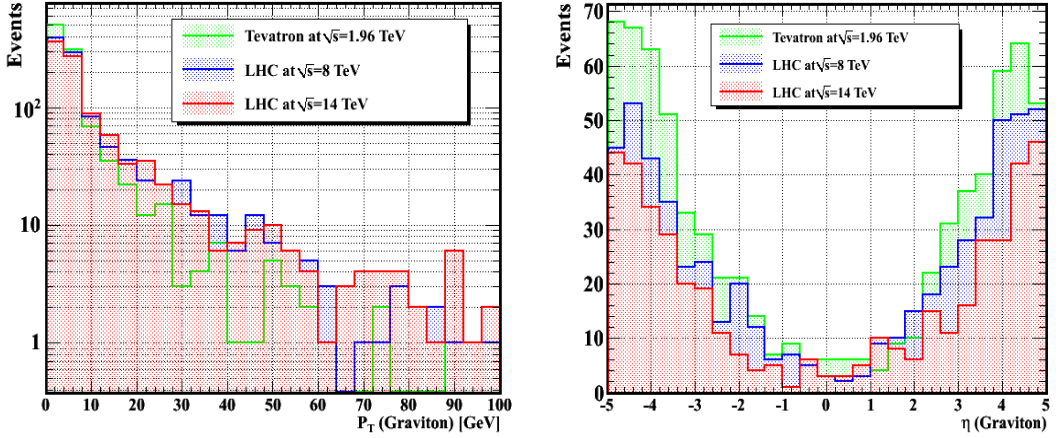

FIG. 7: Transverse momentum (LEFT) and Pesudorapidity (RIGHT) of Graviton in RS model at the LHC $\sqrt{s} = 14$ and 8 TeV with $\Lambda = 5\,TeV$, Graviton mass = 1TeV and at the Tevatron $\sqrt{s} = 1.96\,TeV$ with $\Lambda = 1\,TeV$ and Graviton mass = 0.5 TeV using Pythia8.

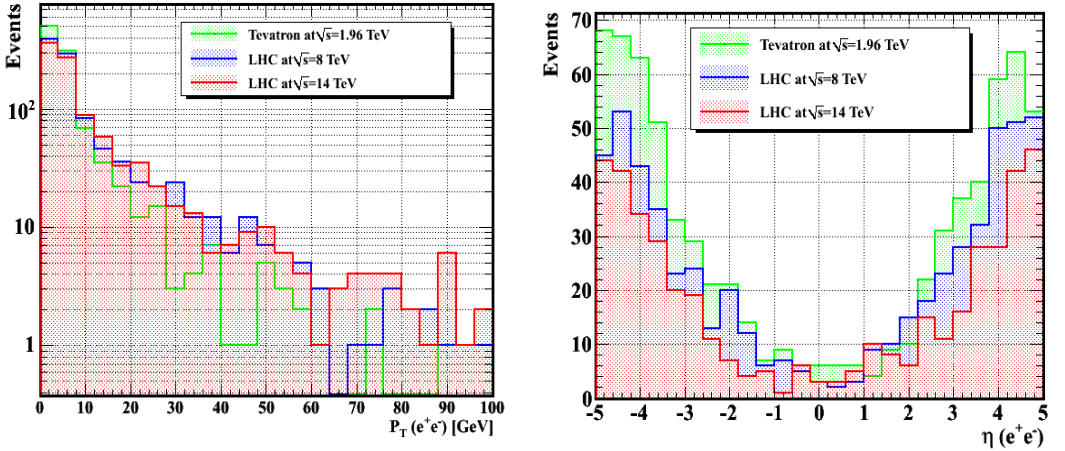

FIG. 8: Transverse momentum (LEFT) and Pesudorapidity (RIGHT) of electrons in RS model at the LHC $\sqrt{s} = 14$ and 8 TeV with $\Lambda = 5\,TeV$, Graviton mass = 1TeV and at the Tevatron $\sqrt{s} = 1.96\,TeV$ with $\Lambda = 1\,TeV$ and Graviton mass = 0.5 TeV using Pythia8.

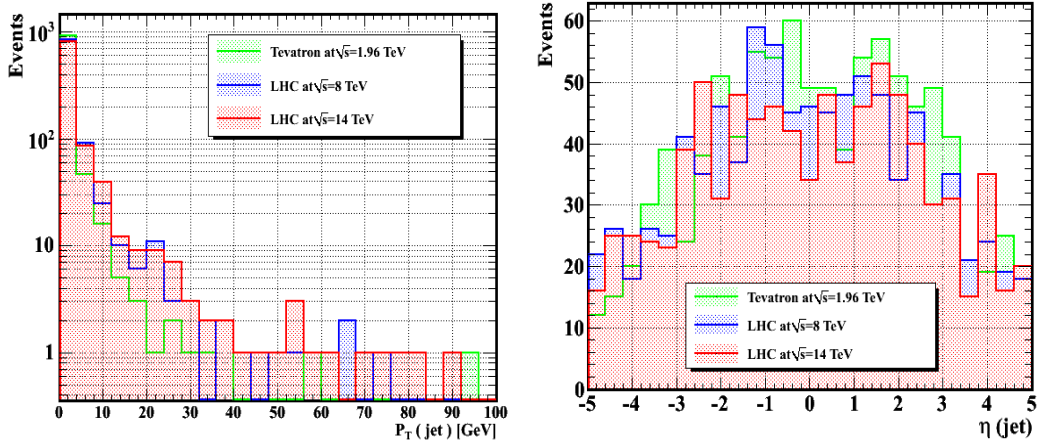

FIG. 9: Transverse momentum (LEFT) and Pesudorapidity (RIGHT) of jet in RS model at the LHC $\sqrt{s} = 14$ and 8 TeV with $\Lambda = 5\,TeV$, Graviton mass = 1TeV and at the Tevatron $\sqrt{s} = 1.96\,TeV$ with $\Lambda = 1\,TeV$ and Graviton mass = 0.5 TeV using Pythia8.

## 4 Conclusions

In this work, we present an analysis for new physics signatures (spin-2 Graviton) in the context of two Large Extra Dimensions (LED) Models using Monte Carlo data. The first model is the ADD model with total dimensions $D = 4 + \delta$ , $\delta = 1,2,3,4,5\ and\ 6$ where $\delta$ is the extra special dimension. From our simulation for ADD model, the Graviton signal can be detected at the LHC (with $\Lambda = 5\,TeV$) and the Tevatron (with $\Lambda = 1\,TeV$) as a missing transverse momentum $P_T^{miss}$ in association with jet in the final state via the process $pp(\bar{p}) \rightarrow G + jet$. The second model is the RS model with 5-dimensions via the process $pp(\bar{p}) \rightarrow G + jet$ , $G \rightarrow e^+e^-$ with final state $e^+e^-$ + jet and the Graviton signal can be detected at the LHC and the Tevatron via high clean signal of two high energy electrons comes from the decay of the spin-2 Graviton.

## AKNOWLEDMENT

It is a pleasure to thank Prof. Torbjorn Sjostrand , Department of Theoretical Physics, Lund University, Lund, Sweden, the main author of Monte Carlo Event Generator (MCEG) Pythia8 for useful discussions. The work by Maxim Yu. Khlopov was supported by the Ministry of Education and Science of Russian Federation project 3.472.2014/K and grant RFBR 14-22-03048.